\documentstyle[12pt]{article}
\input BoxedEPS
\SetRokickiEPSFSpecial 
\HideDisplacementBoxes 

\font\sf=cmss10                    

\begin{document}
\setlength{\baselineskip}{12pt}

\hfill CTP \# 2562

\hfill August, 1996

\vskip 0.20in
\centerline{\large Variational Thomas-Fermi Theory of a 
Nonuniform} 
\vskip 0.05in
\centerline{\large Bose Condensate at Zero Temperature}
\centerline{}
\vskip 0.15in
\centerline{Eddy Timmermans and Paolo Tommasini}
\vskip 0.10in
\centerline{Institute for Theoretical Atomic and Molecular Physics}
\centerline{Harvard-Smithsonian Center for Astrophysics}
\centerline{Cambridge, MA 02138}
\vskip 0.15in
\centerline{Kerson Huang}
\vskip 0.10in
\centerline{Center for Theoretical Physics, Laboratory for
Nuclear Science}
\centerline{       and Department of Physics}
\centerline{   Massachusetts Institute of Technology}
\centerline{   Cambridge, MA 02139}

\vskip 0.60in

\centerline{                ABSTRACT}
\vskip 0.20in

We derive a description of the spatially inhomogeneous Bose-Einstein 
condensate which treats the system locally as a homogeneous system.
This approach, similar to the Thomas-Fermi model for the inhomogeneous
many-particle fermion system, is well-suited to describe the atomic
Bose-Einstein condensates that have recently been obtained
experimentally through atomic trapping and cooling.
In this paper, we confine our attention to the zero temperature case, 
although the treatment can be generalized to finite temperatures, as we shall
discuss elsewhere.

Several features of this approach, which we shall call the 
Thomas-Fermi-Bogolubov description, are very attractive:
1.  It is simpler than the Hartree-Fock-Bogolubov technique.
We can obtain analytical results 
in the case of weakly interacting bosons
for quantities such as the chemical potential,
the local depletion, pairing, pressure and density of states.
2.  The method provides an estimate for the error due to the inhomogeneity
of the bose-condensed system. This error is a local quantity so that 
the validity of the description for a given trap and a given number of trapped
atoms, can be tested as a function of position.
We see for example that at the edge of the 
condensate, the Thomas-Fermi-Bogolubov theory always breaks down. 
3. The Thomas-Fermi-Bogolubov
description can be generalized to
treat the statistical mechanics of the bose gas at finite
temperatures.

\newpage
\section{Introduction}

The recently reported Bose-Einstein condensates of trapped neutral atoms
\cite{AE} -- \cite{KM} represent the first unambiguous 
observations of a weakly interacting bose condensed gas.  
Quantitatively, we can characterize the strength
of the interaction by the expansion parameter in the perturbation
treatment of the homogeneous bose gas , $\sqrt{n a^{3}}$,
where $n$ is the density and $a$ the scattering length of the inter-atomic 
potential.  In the atomic-trap experiments, typically n $\sim 10^{12}$ 
$-$  $10^{14} \rm{cm}^{-3}$,
and a $\sim 1-5 $ nm, so that  $\sqrt{n a^{3}} \sim 10^{-2}$$\; - \; 
$ $3 \times 10^{-5}$.  
Thus, in the sense of perturbation theory, the
observed condensates are indeed textbook examples of weakly interacting
systems. 
For the uniform bose gas, perturbation theory leads to simple analytical
results.  Although the trapped condensates can be described by means of
the Hartree-Fock Bogolubov equations \cite{GG}-\cite{HP},
the latter approach does not lend itself to an analytical perturbation
treatment. 

Intuitively, one expects that a many-body system whose density varies 
slowly in space can be described locally as a homogeneous system.
Based on this picture, the Thomas-Fermi method 
\cite{F} \cite{T} was proposed
for the calculation of the electron density in a heavy atom.  Lieb and Simon
\cite{LS} showed that the treatment is exact in the limit when the atomic
number goes to infinity.  Application to a confined Bose condensate was
pioneered by Goldman, Silvera and Legget \cite{GSL}, and recently
reconsidered by Chou, Yang, and Yu \cite{CN}-\cite{CL}.
As pointed out by Kagan, Shlyapnikov, and Walraven\cite{KSW}, 
the local-density description is valid when
\begin{equation}
\mu/\hbar\omega\gg 1 \; ,
\end{equation}
where $\mu$ is the mean-field energy per particle, or chemical potential,
and $\hbar\omega$ is the zeropoint energy in the trap.

In this paper, we derive such a description from first principles within 
the framework of the variational technique. 
We emphasize that, unlike the practice of neglecting the
kinetic energy term in the Gross-Pitaevski equation which in the recent
literature is sometimes called the Thomas-Fermi approximation, 
the resulting variational description is not limited to the condensate, but
describes the depletion, pressure and all other thermodynamic quantities.
Furthermore, like the uniform gas, the Thomas-Fermi theory leads to a
perturbation treatment of the weakly interacting condensates, giving
simple analytical expressions for these quantities.
Another important advantage of the
Thomas Fermi treatment is that it can be generalized to describe
finite temperature systems, as we shall discuss 
in future work.  In this paper we focus on the bose gas at zero temperature. 

The paper is organized as follows.  In section 2, we generalize the usual
Bogolubov transformation to describe spatially inhomogeneous condensates.
In section 3, we introduce the Wigner representation and gradient
expansion, which provide the tools to describe the nearly homogeneous
systems and make the Thomas-Fermi approximation.  The advantage of this
systematic approach to the Thomas-Fermi approximation is that it
provides an estimate of the error incurred by the inhomogeneity of
the condensate, allowing one to estimate the accuracy of the Thomas
Fermi results.
We consider this point to be very important in
view of the 
fact that some traps, depending on the potential and 
the number of trapped
atoms, are too far from homogeneity to be described by a Thomas-Fermi 
description.  In addition, even if the Thomas-Fermi
description is valid in the middle of the trap, it 
breaks down at the edge of the condensate.
In sections 4 and 5, we obtain the mean-field description of the bose 
system in the Thomas-Fermi approximation.  The equations, derived within
the framework of the variational principle, provide a fully self-consistent
description, indicating that the Thomas-Fermi decription is by no means
limited to weakly interacting systems.  
This remark can be expected to be of future importance
in the light of recent experimental efforts to obtain condensates of
higher density.
Nonetheless, because of the special interest in the weakly interacting
systems, we proceed in section 6 to derive a perturbation treatment and 
obtain analytical 
results for quantities
such as the chemical potential, the local depletion, pairing and pressure.
With the experimental atomic-traps in mind,
we apply the results of the general theory to the special case
of a trapping potential that is of the type of
a simple spherically symmetric harmonic oscillator in section 7.
Finally, in section 8, we derive a density of states of the
trapped weakly interacting condensate within the spirit of the 
Thomas-Fermi approximation.

\section{Generalized Bogolubov Transformation}

The Bogolubov quasi-particle concept \cite{NP} provides a very elegant
description of the interacting Bose-Einstein condensate. 
The quasi-particles are represented by 
creation ($\eta^{\dagger}$) and annihilation ($\eta$) operators
that are linear combinations of regular 
single-particle creation ($a^{\dagger}$) and annihilation
($a$) operators. In treating a homogeneous system, for which we can work
in a  basis of single-particle plane-wave states of momentum ${\bf k}$, 
the Bogolubov transformation which relates the quasi-particle
and regular particular operators, takes on a particularly simple form,
\begin{eqnarray}
\eta^{\dagger}_{\bf k} &=& x_{\bf k} a^{\dagger}_{\bf k} +
y_{\bf k} a_{-{\bf k}} \; , \nonumber \\
\nonumber \\
\eta_{\bf k} &=& x_{\bf k} a_{\bf k} + y_{\bf k} a^{\dagger}_{-{\bf k}} \; ,
\label{e:BT}
\end{eqnarray}
where for the purpose of describing the static properties of a 
condensate in equilibrium, we can limit the transformation 
parameters, $x_{\bf k},y_{\bf k}$ 
to real numbers.  
Furthermore, the isotropy of the many-body system suggests
that the transformation parameters only depend on the magnitude
of the momentum, $x_{\rm{\bf k}}$ = $x_{\rm{k}}$ and $y_{\rm{\bf k}}$ =
$y_{\rm{k}}$.
Requiring the quasi-particle operators to be canonical,
$\left[ \eta_{\bf k},\eta^{\dagger}_{{\bf k}'} \right]=
\delta_{{\bf k},{{\bf k}'}}, 
\left[ \eta_{\bf k},\eta_{\bf k'} \right] =
\left[ \eta^{\dagger}_{\bf k}, \eta^{\dagger}_{{\bf k}'} \right] = 0$,
gives an additional constraint to $x_{\rm{k}}$ and 
$y_{\rm{k}}$,
\begin{equation}
x^{2}_{k} - y^{2}_{k} = 1 ,
\label{e:cans}
\end{equation}
from which we can see that a single parameter $\sigma_{k}$, 
with $x_{k}$ = $\cosh\sigma_{k}$ and $y_{k}$ = $\sinh\sigma_{k}$, 
suffices to parametrize the Bogolubov transformation (\ref{e:BT}).
In addition, with Eq.(\ref{e:cans}),
we can also write the
Bogolubov transformation as 
\begin{eqnarray}
a^{\dagger}_{\bf k} &=& x_{k} \eta^{\dagger}_{\bf k} - 
y_{k} \eta_{-{\bf k}} \; , \nonumber \\
\nonumber \\
a_{\bf k} &=& x_{k} \eta_{\bf k} - y_{k} \eta^{\dagger}_{-{\bf k}} \; ,
\label{e:BTI}
\end{eqnarray}
which is the inverse transformation of (\ref{e:BT}).

It is useful to define the following quantities, the ``distribution 
function'' $\rho$ and the ``pairing function'' $\Delta$:
\begin{eqnarray}
\rho_{\bf k} &=& \langle a_{\bf k}^{\dagger} a_{\bf k} \rangle
= \frac{1}{2} \left[ \cosh 2 \sigma_{\bf k} -1 \right] \; , \nonumber\\
\Delta_{\bf k} &=& -  \langle a_{\bf k} a_{-\bf k} \rangle
= \frac{1}{2} \sinh 2 \sigma_{\bf k} \; ,
\end{eqnarray}
where the brackets $\langle \; \rangle$ represent the ground state 
expectation value.
The best values for $x_{\rm{k}}$ and $y_{\rm{k}}$ are obtained 
variationally by minimizing the ground state free energy.

As stated, the above description (1)-(3) 
only applies to homogeneous
systems, whereas the treatment of a general (inhomogeneous) 
condensate, as we shall
show below, involves a Bogolubov transformation that is
quite
different in appearance, from the homogeneous case.  
However, we can expect the results of the
homogeneous treatment to describe the
`local' behavior of an inhomogeneous condensate, provided the
spatial variations of the condensate are sufficiently slow.
In describing many-particle Fermion systems, this intuitive 
picture forms one of the key ingredients of the well-known 
Thomas-Fermi description of slowly varying many-particle systems.

To arrive at a general treatment, we choose to work with
boson-field operators, $\hat{\Psi}({\bf x})$ and  
$\hat{\Psi}^\dagger({\bf x})$, an approach that 
offers the advantage of not having to specify a basis a priori. Furthermore,
in the presence of a condensate, it is convenient to work
with the fields $\hat{\psi}({\bf x})$ and 
$\hat{\psi}^{\dagger}({\bf x})$, which are displaced from
the original fields $\hat{\Psi}({\bf x})$ and $\hat{\Psi}^{\dagger}({\bf x})$
by the expectation value $\phi({\bf x})$ of $\hat{\Psi}({\bf x})$,
\begin{eqnarray}
\hat{\Psi}({\bf x}) &=& \hat{\psi}({\bf x}) + \phi({\bf x}),
\nonumber  \\
\hat{\Psi}^{\dagger}({\bf x}) &=& \hat{\psi}^{\dagger}({\bf x}) + 
\phi^{\ast}({\bf x}),
\end{eqnarray}
where, for the purpose of describing the static properties of a condensate in
equilibrium,
$\phi$ can be taken to be real, and where
$\hat{\psi}({\bf x})$ and $\hat{\psi}^{\dagger} ({\bf x})$ are the 
displaced field operators which satisfy the canonical commutation
relation, $\left[ \hat{\psi}({\bf x}), \hat{\psi}^{\dagger} ({\bf x}') 
\right] $ $= \delta ({\bf x}-{\bf x}')$, and furthermore, 
\begin{equation}
\langle \hat{\psi}({\bf x}) \rangle  = \langle \hat{\psi}^{\dagger} 
({\bf x}) \rangle = 0.
\label{e:expo}
\end{equation}

We introduce the Bogolubov transformation as a general 
linear transformation relating the displaced fields to the
quasi-particle fields, $\hat{\xi}({\bf x})$
and $\hat{\xi}^{\dagger}({\bf x})$,
\begin{eqnarray}
\hat{\psi}({\bf x}) &=& \int d^{3}z \left[X({\bf x},{\bf z})\hat{\xi}({\bf z})
-Y({\bf x},{\bf z})\hat{\xi}^{\dagger}({\bf z})\right] , \nonumber \\
\nonumber\\
\hat{\psi}^{\dagger}(\bf x) &=& \int d^{3}z \left[X^{\ast}({\bf x}
,{\bf z})\hat{\xi}^{\dagger}({\bf z})
-Y^{\ast}({\bf x},{\bf z})\hat{\xi}({\bf z})\right], 
\label{e:GBT}
\end{eqnarray}
which is the generalization of Eq.(\ref{e:cans}).
The non-local nature of the generalized Bogolubov transformation
(\ref{e:GBT}) should not be surprising $-$ the `homogeneous' Bogolubov
transformation (\ref{e:BT}) can be written in the same form
with the special feature, due
to the homogeneity of the system, that $X({\bf x}, {\bf y})$ and $Y({\bf x},
{\bf y})$ only depend on ${\bf x} 
- {\bf y}$. 

Requiring the quasi-particle fields to be canonical,
leads to
\begin{equation}
\int d^{3} z \left[ X({\bf x},{\bf z}) X({\bf z},{\bf y}) - 
Y({\bf x},{\bf z})
Y({\bf z},{\bf y})\right]  = \delta({\bf x}-{\bf y}) \; ,
\label{e:cang}
\end{equation}
which is the generalization of (\ref{e:cans}).

It is possible to derive equations for the inhomogeneous bose systems by
variationally determining the best transformations $X$ and $Y$,
minimizing the free energy.  This however, is not the path we
choose to follow here. Instead, we manipulate the generalized
Bogolubov transformations in a manner similar to the procedure
to obtain the Wigner distribution from the off-diagonal
single-particle density function.  Once this is achieved, the 
steps that lead to a Thomas-Fermi description are known from 
quantum transport theory.  One interesting new aspect of this
treatment is that the central object of the theory 
is not a distribution 
function, which in some sense can still  be regarded as an observable,
but a transformation.
Although this transformation determines the value of all observables,
it is clearly not an observable quantity by itself.

\section{Wigner Representation and Gradient Expansion}

Wigner showed that a quantum mechanical single-particle system,
costumarily characterized by its wave function 
$ \Psi ({\bf x}) $,
can alternatively be fully characterized by a different function,
\begin{equation}
\rho_{\rm{W}}({\bf R},{\bf p}) = 
\int d^{3} r \;  
\Psi^{\ast}({\bf R}+{\bf r}/2) \Psi({\bf R}-
{\bf r}/2)
\exp (i{\bf p}\cdot{\bf r}) ,
\label{e:WR}
\end{equation}
where here $-$ as in the rest of the paper $-$ we work in units in 
which $\hbar$=1. This function (\ref{e:WR}), 
known as the Wigner Distribution function,
can be interpreted as a phase space distribution function \cite{WG} and
leads to a description that is remarkably close to classical mechanics.
The analogy with a classical phase space distribution function
is not complete (for example $ \rho_{\rm W} $ can take on
negative values), but can be justified by the fact that the quantum mechanical
expectation value of observables are 
equal to the `phase space
integrals' of the corresponding classical quantities, weighted by
$(2 \pi)^{-3} \rho_{{\rm W}} $,
\begin{eqnarray}
\langle \Psi | f | \Psi \rangle & = &
\int d^{3} x \; \Psi^{\ast}({\bf x})
f({\bf x}) \Psi({\bf x})\nonumber \\ 
& = & (2 \pi)^{-3}
\int d^{3}  p \; \int d^{3} R \; f({\bf R}) \; 
\rho_{\rm{W}}({\bf R},{\bf p}),
\nonumber\\
\langle \Psi |\hat{{\bf p}} | \Psi \rangle & = &
\int d^{3} x \; \Psi^{\ast}({\bf x})
\hat{{\bf p}} \Psi({\bf x}) \nonumber\\
& = & (2 \pi)^{-3}
\int d^{3} p \; \int d^{3} R \; {\bf p} \; 
\rho_{\rm{W}}({\bf R},{\bf p}), 
\end{eqnarray}
etc. More recently, the many-particle generalization of the Wigner
distribution has found many important applications in diverse areas such
as nuclear \cite{DL} and solid state physics \cite{DD}.

An important motivation to work in the transformed
representation of Eqs.(\ref{e:WR}), 
$({\bf x},{\bf x}') \rightarrow ({\bf R},{\bf p})$,
\begin{equation}
A_{W}({\bf R},{\bf p}) = \int d^{3} r \;
A({\bf R} + {\bf r}/2, {\bf R} - {\bf r}/2)
\exp (i{\bf p}\cdot{\bf r}) ,
\label{e:wrep}
\end{equation}
and its inverse
\begin{equation}
A({\bf x} , {\bf x}') =
(2 \pi)^{-3} \int d^{3} p \; A_{W} (\left[ {\bf x}+{\bf x}' \right] /2,{\bf p}) 
\; \exp(-i{\bf p}\cdot\left[ {\bf x}-{\bf x}' \right]),
\label{e:wrepi}
\end{equation}
which shall henceforth be referred to as the Wigner representation, is
that it is extraordinarily well suited to describe
nearly homogeneous systems.  This convenient feature  
follows from the gradient expansion \cite{DL} -- \cite{DD}.  
The gradient expansion
shows that, to first
order, a `product' operator $C({\bf x},{\bf x}') = \int
d^{3} z \; A({\bf x},{\bf z}) B({\bf z},{\bf x}') $, in
the Wigner representation simply gives the algebraic product of A
and B, $C_{W}({\bf R},{\bf p}) \approx A_{W}({\bf R},{\bf p}) 
B_{W}({\bf R},{\bf p})$.  The higher-order corrections
to this approximation can be written as a series of
terms containing successively higher-order derivatives
in the (${\bf R},{\bf p}$)$-$ coordinates,
\begin{eqnarray}
C_{W}({\bf R},{\bf p}) \approx
A_{W}({\bf R},{\bf p}) B_{W}({\bf R},{\bf p}) + \frac{1}{2 \rm{i}} 
\sum_{j=1}^{3} \left[ \frac{\partial A_{W}}{\partial R_{j}}
\frac{\partial B_{W}}{\partial p_{j}} - \frac{\partial A_{W}}{\partial p_{j}}
\frac{\partial B_{W}}{\partial R_{j}} \right]\nonumber\\
- \frac{1}{8}  \sum_{j=1}^{3} \left[
\frac{\partial^{2} A_{W}}{\partial R_{j}^{2}}
\frac{\partial^{2} B_{W}}{\partial p_{j}^{2}}
+ \frac{\partial^{2} A_{W}}{\partial p_{j}^{2}}
\frac{\partial^{2} B_{W}}{\partial R_{j}^{2}}
-  2 \frac{\partial^{2} A_{W}}{\partial R_{j} \partial p_{j}}
\frac{\partial^{2} B_{W}}{\partial R_{j} \partial p_{j}} \right] + \cdots 
\; .
\label{e:gradexp}
\end{eqnarray}
The first order correction in the gradient expansion (\ref{e:gradexp})
is $\left\{ A_{W},B_{W} \right\}_{\rm{{PB}}}$,
the Poisson bracket of $A_{W}$ and $B_{W}$. If we  know that the range 
of $A_{W}$ and $B_{W}$ in ${\bf p}$-space is of the order of $p_{c}$, then 
the magnitude of the derivatives 
$\partial B_{W} / \partial p $ and $\partial^{2} B_{W} / \partial p^{2}$
in (\ref{e:gradexp}) can
be estimated to be of the order of $B_{W}/p_{c}$ and $B_{W}
/p_{c}^{2}$ respectively.
This approximation will allow us to obtain a very simple estimate of
the `inhomogeneity' 
error.

At this point, we return to the generalized Bogolubov transformation,
$X({\bf x},{\bf y}),Y({\bf x},{\bf y})$, of the previous section.
Working in the Wigner representation and expanding the `canonicity'
relation (\ref{e:cang})
between between $X$ and $Y$ in the manner of the gradient
expansion, we find up to first order in the spatial derivatives,
a relation that is similar to the constraint equation (\ref{e:cans})
of the homogeneous Bogolubov transformation,
\begin{equation}
X^{2}_{W}({\bf R},{\bf p}) - Y^{2}_{W}({\bf R},{\bf p}) \approx 1 .
\label{e:canw}
\end{equation}
Consequently, the general Bogolubov transform can be parametrized in the
same way as the Bogolubov transform for the homogeneous bose gas,
$X_{W}({\bf R},{\bf p})$ = $\cosh[\sigma ({\bf R},{\bf p})]$,
$Y_{W}({\bf R},{\bf p})$ = $\sinh[\sigma ({\bf R},{\bf p})]$,
where for the slowly varying condensate, 
the $\sigma$ $-$ parameters depend on the momentum {\sl and}
position : $\sigma ({\bf R},{\bf p})$.
The distribution and pairing functions,
$\rho({\bf{x}},{\bf{x}}')$ = $\langle \hat{\psi}^{\dagger} (\bf{x}) \hat{\psi} 
(\bf{x}') \rangle $ and $\Delta({\bf{x}},{\bf{x}}')$ 
= $ - \langle \hat{\psi} (\bf{x}) 
\hat{\psi}
(\bf{x}') \rangle$ take on the following form in the Wigner representation: 
\begin{eqnarray}
\rho_{W} ({\bf R},{\bf p}) &=&
\frac{1}{2} \left[ \cosh (2 \sigma ({\bf R},{\bf p})) -1 \right] ,
\nonumber\\
\Delta_{W} ({\bf R},{\bf p}) &=&
\frac{1}{2} \sinh (2 \sigma ({\bf R},{\bf p})) .
\label{e:rds1}
\end{eqnarray}
The local $\sigma$ parametrization of the Bogolubov transformation
is crucial to the Thomas-Fermi description and it is
upon the validity of (\ref{e:canw}) that the Thomas-Fermi theory rests.
The error introduced to (\ref{e:canw}) due to the inhomogeneity of
the system can be estimated by the lowest order
non-vanishing term in the gradient expansion (\ref{e:gradexp}).  
Notice that the first order term in the gradient expansion
of (\ref{e:canw}) vanishes since it is the 
the sum of Poisson brackets of quantities with themselves.
Consequently, the error has to be estimated from the second order term.

\section{Energy Density}

In the variational method, the quantity to minimize
is F, the ground state 
free energy, which we can put in a `local' form,
$F = \int d^{3} R \; f({\bf R})$, where 
$f({\bf R})$ is 
the energy density.  We achieve this
result in two steps.  In the first step, we shift to the Wigner
representation in the integrand for the mean field expression
for the ground state
energy.  In the second step, we notice that the short-range
nature of the inter-atomic interaction renders the resulting
integrand essentially `local', i.e. the integrand contains
only $\sl single$ (not double) integrals over the position
variables.

The ground state free energy is the 
expectation value of $\hat{H} - \mu \hat{N}$, where 
$\hat{H}$ is the many-body hamiltonian of the boson
system, $\hat{N}$, the number operator and $\mu$, the
chemical potential:  
\begin{equation}
{\hat{H}} - \mu \hat{N}
= \int d^3 x \; \hat{\Psi}({\bf x})^{\dag} \hat{h}({\bf x}) 
\hat{\Psi}({\bf x})  + 
\frac{1}{2}\int d^3 x \; d^3y \;
\hat{\Psi}({\bf y})^{\dag}\hat{\Psi}({\bf x})^{\dag} V(|{\bf x}-{\bf y}|)
\hat{\Psi}({\bf x}) \hat{\Psi}({\bf y}),
\label{e:ham}
\end{equation}
where $V(|{\bf x}-{\bf y}|)$ represents the inter-atomic potential
and $\hat{h}({\bf x})$ is the one-body part of the free energy, 
\begin{equation}
\hat{h}({\bf x}) = -\frac{\nabla^{2}}{2 m} + V_{\rm ext}({\bf x}) -\mu,
\end{equation}
where $V_{\rm ext}({\bf x})$ is the external potential.

The presence of a condensate displaces
the field operators ${\hat{\Psi}} ({\bf x})$ by their expectation 
value $\phi ({\bf x})$. 
To generate the variational free energy, we shall use 
the mean field approximation, in which terms of first and
third order in $\hat{\psi}$ and $\hat{\psi}^{\dagger}$ vanish
(\ref{e:expo}) and the fourth order term factorizes as follows: 
\begin{eqnarray}
&& \langle \hat{\psi}({\bf y})^{\dag}\hat{\psi}({\bf x})^{\dag}
\hat{\psi}({\bf x}) \hat{\psi}({\bf y}) \rangle 
\approx \nonumber\\
\nonumber\\
&& \Delta ^{\ast}({\bf y},{\bf x})  \Delta ({\bf y},{\bf x})
+ \rho ({\bf y},{\bf x})  \rho ({\bf x},{\bf y})
+ \rho ({\bf x},{\bf x})  \rho ({\bf y},{\bf y}) .
\end{eqnarray}
The variational nature of this procedure is insured by
the existence of a variational ground state that gives this type of
factorization.  In fact, the variational ground state corresponds to the choice
of the gaussian wave functional \cite{HP}. 

The displacement of the fields and the factorization of the 
expectation values, although straightforward, gives rise to
a somewhat lengthy expression for the free energy.  It is then
convenient to classify the different contributions by their
order in $\phi$ and their functional dependence on $\rho$ and
$\Delta$:

\noindent
1. $h_{1}$ is the one-body contribution of zeroth
order in $\phi$ to the ground state energy,
\begin{equation}
h_{1} = \frac{1}{2} \int d^{3}x \; d^{3}y \;
\hat{h}({\bf x}) \rho ({\bf y},{\bf x}) 
\delta ({\bf x}-{\bf y}) .
\end{equation}
2.  In analogy with the Hartree-Fock theory, we call $ V_{\rm{dir}} $ 
given below, the direct energy contribution to the energy, 
\begin{equation}
V_{\rm{dir}} = \frac{1}{2} \int d^{3}x \; d^{3}y \; 
\rho ({\bf y},{\bf y})  \rho ({\bf x},{\bf x}) V(|{\bf x}-{\bf y}|).
\end{equation}
3.  Using the same analogy to the Hartree-Fock treatment, the 
exchange energy, $V_{\rm{exch}} $, is equal to 
\begin{equation}
V_{\rm{exch}} = \frac{1}{2} \int d^{3}x \; d^{3}y \;
\rho ({\bf x},{\bf y})  \rho ({\bf y},{\bf x}) V(|{\bf x}-{\bf y}|).
\end{equation}
4. Standard Hartree-Fock theory does not describe pairing and the pairing
energy,  $ V_{\rm pair} $, 
\begin{equation}
V_{\rm{pair}} = \frac{1}{2}  \int d^{3}x \; d^{3}y \; 
\Delta^{\ast} ({\bf y},{\bf x})  \Delta ({\bf y},{\bf x}) 
V(|{\bf x}-{\bf y}|) \; ,
\end{equation}
is consequently absent from
the Hartree-Fock expressions.

In second order in $\phi$, we find contributions
that can be
obtained from the above terms by replacing either
$\Delta ({\bf x},{\bf y})$ or $\rho ({\bf x},{\bf y})$
by $\phi ({\bf x}) \phi ({\bf y})$. 

\noindent
5. For example,
the one-body contribution, due to the kinetic and potential energy
of the condensate is $h_{1}^{\phi}$, where
\begin{equation}
h_{1}^{\phi} = \int d^{3}x \; \phi ({\bf x})
\hat{h}({\bf x}) \phi ({\bf x}) .
\end{equation}
6.  $ V_{\rm{dir}}^{\phi} $ is the direct contribution
to the interaction energy, stemming from the interaction of the condensate with
the particles that have been `forced' out of the condensate (depletion), 
\begin{equation}
V_{\rm{dir}}^{\phi} = \frac{1}{2}  \int d^{3}x \; d^{3}y \;
\phi ({\bf y}) \phi ({\bf y})  \rho ({\bf x},{\bf x}) V(|{\bf x}-{\bf y}|).
\end{equation}
7.  Similarly, $ V_{\rm{exch}}^{\phi} $ 
is the exchange contribution of
second order in $\phi$,
\begin{equation}
V_{\rm{exch}}^{\phi} = \frac{1}{2}  \int d^{3}x \; d^{3}y \;
\phi ({\bf y}) \phi ({\bf x})  \rho ({\bf y},{\bf x}) V(|{\bf x}-{\bf y}|).
\end{equation}
8.  We represent the pairing energy of the condensate with the particles out
of the condensate by $ V_{\rm{pair}}^{\phi} $, 
\begin{equation}
V_{\rm{pair}}^{\phi} = \frac{1}{2}  \int d^{3}x \; d^{3}y \;
\phi ({\bf y}) \phi ({\bf x})  \Delta ({\bf y},{\bf x}) V(|{\bf x}-{\bf y}|).
\end{equation}
9.  Finally, we denote the contribution of fourth order in $\phi$,
representing the interaction energy of the condensate with
itself, by $ V^{\phi\phi} $ :
\begin{equation}
V^{\phi\phi} = \frac{1}{2}  \int d^{3}x \; d^{3}y \;
\phi^{2} ({\bf y}) \phi^{2} ({\bf x})  V(|{\bf x}-{\bf y}|).
\end{equation}
With this notation, the mean-field expression for the ground state energy reads
\begin{eqnarray}
F &=& \langle \hat{H} - \mu \hat{N} \rangle \nonumber\\
&=& h_{1} +
V_{\rm{dir}} +
V_{\rm{exch}} +
V_{\rm{pair}} + \nonumber\\
&& h_{1}^{\phi} + 2
V_{\rm{dir}}^{\phi} + 2
V_{\rm{exch}}^{\phi} - 2
V_{\rm{pair}}^{\phi} +
V^{\phi\phi} ,
\end{eqnarray}
where the minus sign of the $V_{pair}^{\phi}$ term stems from the definition
of $\Delta$ = $- \langle \hat{\psi} \hat{\psi} \rangle$.

At this point, we introduce the Wigner representation into the integrands
of the above contributions to the mean-field expressions for the ground
state free energy. The resulting expressions 
resemble the corresponding terms 
for the homogeneous gas, with an additional label ${\bf R}$ over which
is integrated.  
For the sake of notational convenience we introduce the following 
integration symbol $\int_{\bf R}$ or $\int_{\bf p}$, which
represents the usual integral over all of space, $\int d^{3} {\bf R}$,
if ${\bf R}$ 
is a position variable or
$(2\pi)^{-3} \int d^{3} p $, if ${\bf p}$ is a momentum variable: 
\begin{eqnarray}
\int_{\bf p} &\equiv& (2\pi)^{-3} \; \int d^{3} p \; ,
\nonumber\\
\int_{\bf R} &\equiv& \int d^{3} R \; .
\end{eqnarray}
The terms of zero order in $\phi$ then give 
\begin{eqnarray}
h_{1}
&=& 
\int_{\bf R} \; \int_{\bf p} \left[ \frac{p^{2}}{2m} + V({\bf R}) - \mu
\right] \rho_{W} ({\bf R},{\bf p}), \nonumber\\
\nonumber\\
V_{\rm{exch}} &=& 
\int_{\bf R} \; \int_{\bf p} \; \int_{{\bf p}'} \; \rho_{W} ({\bf R},{\bf p})
v({\bf p}-{\bf p}') \rho_{W} ({\bf R},{\bf p}') \; ,\nonumber\\
V_{\rm{pair}} &=& 
\int_{\bf R} \; \int_{\bf p} \; \int_{{\bf p}'} \; \Delta_{W} ({\bf R},{\bf p})
v({\bf p}-{\bf p}') \Delta_{W} ({\bf R},{\bf p}') \; , 
\nonumber\\
V_{\rm{dir}} &=&
\int_{\bf R} \; \int_{\bf r} \; \int_{\bf p} \; \int_{{\bf p}'} \; \int_{\bf q}
\rho_{W} ({\bf R} - {\bf r}/2 ,{\bf p} ) 
\rho_{W} ({\bf R} + {\bf r}/2 ,{\bf p}')
\exp (i {\bf q}\cdot{\bf r})
v({\bf q})\; ,
\end{eqnarray}
where $v$ is the Fourier transform of the interaction potential,
$v({\bf q})$ = $\int d^{3} r V({\bf r}) \exp (-i{\bf q}\cdot{\bf r})$.

The terms that are of second order in $\phi$ can be obtained
by replacing one $\rho$ or $\Delta$ by $\phi \phi$.  In the Wigner 
representation, this procedure yields expressions that are similar
to the corresponding terms of zero order in $\phi$ with 
$\rho_{W} ({\bf R},{\bf p})$ or $\Delta_{W} ({\bf R},{\bf p})$ replaced
by a function $Q_{W}({\bf R},{\bf p})$, where
\begin{equation} 
Q_{W}({\bf R},{\bf p}) = 
\int_{\bf r} \; \phi ({\bf R} + {\bf r}/2) 
\phi ({\bf R} - {\bf r}/2) \exp (i{\bf p}\cdot{\bf r}) .
\end{equation}
Notice that the contributions of second order in $\phi$, 
are non-local in the sense that
their expressions contain integrals over more than one position
variable.  Nevertheless, if we consider the scale  
on which the physical quantities vary in space,
or in momentum space, it becomes apparent that 
the non-local integrals can be approximated by local expressions.
We illustrate this point by considering the exchange 
($V_{exch}^{\phi}$) and pairing ($V_{pair}^{\phi}$) energies.
The key to obtain local expressions is to notice that $Q_{W}({\bf R},{\bf p})$ 
varies with respect to ${\bf p}$ on the scale of ${\bf R}_{0}^{-1}$, where
${\bf R}_{0}$ is the size
of the condensate.  On the other hand, $v({\bf p}-{\bf p}')$
varies on the scale of $l_{ r} ^{-1}$ where $l_{ r}$ is the
range of the atom-atom interaction.  Typically ${\bf R}_{0} \gg l_{ r}$
so that $Q_{W}({\bf R},{\bf p})$ varies much more rapidly with respect
to ${\bf p}$ than  $v({\bf p}-{\bf p}')$.  In fact, when ${\bf p}$
is large enough to make $v({\bf p}-{\bf p}')$ significantly different
from $v({\bf p}')$,  $Q_{W}({\bf R},{\bf p}) \approx 0$.  Thus, we can
replace  $v({\bf p}-{\bf p}')$ by  $v({\bf p}')$ in the integrands :
\begin{eqnarray}
V_{\rm{exch}}^{\phi} &\approx&
\frac{1}{2}
\int_{\bf R} \int_{\bf r}  \int_{\bf p}  \int_{{\bf p}'} \;
\phi ({\bf R} + {\bf r}/2)
\phi ({\bf R} - {\bf r}/2) \exp (i{\bf p}\cdot{\bf r})
v({\bf p}') \rho_{W} ({\bf R},{\bf p}') 
\nonumber\\
&=& \frac{1}{2}  \phi^{2} ({\bf R})
\int_{\bf R} \int_{{\bf p}'} v({\bf p}') \rho_{W} ({\bf R},{\bf p}'),
\nonumber\\
V_{\rm{pair}}^{\phi} &\approx&
\frac{1}{2} \phi^{2} ({\bf R}) \int_{\bf R}  \int_{{\bf p}'} \; 
v({\bf p}') \rho_{W} ({\bf R},{\bf p}') .
\end{eqnarray}

The same considerations regarding the relative magnitude of the
relevant length scales show that we can similarly simplify the expression
of the $\phi^{4}$ interaction energy, 
$V^{\phi\phi} $, and the direct interaction
energies, $V_{\rm{dir}}$ and 
$V_{\rm{dir}}^{\phi} $.
The local expressions are most easily obtained by considering the difference
in length scales before introducing
the Wigner representation.  In coordinate space, we notice
that $\rho ({\bf x},{\bf x}) \approx \rho ({\bf y},{\bf y})$ if
$|{\bf x}-{\bf y}| \le l_{ r}$.  Thus, we can replace 
$\rho ({\bf x},{\bf x})$ by $\rho ({\bf y},{\bf y})$ in an
integrand if it is accompanied by $V(|{\bf x}-{\bf y}|)$:
\begin{eqnarray}
V_{\rm{dir}} &\approx& \frac{1}{2}\int d^{3} x \; d^{3} y \;
\rho^{2} ({\bf x},{\bf x}) V(|{\bf x}-{\bf y}|) \nonumber\\
&=& \frac{1}{2} v(0) \int_{\bf R} \int_{\bf p} \; \int_{{\bf p}'} \;
\rho_{W} ({\bf R},{\bf p}) \rho_{W} ({\bf R},{\bf p}') \; , \nonumber\\
V_{\rm{dir}}^{\phi} &\approx& \frac{1}{2} v(0) 
\int_{\bf R} 
\int_{\bf p} \; \phi^{2} ({\bf R}) \rho_{W} ({\bf R},{\bf p}) \nonumber\\
V^{\phi\phi} &\approx& \frac{1}{2} v(0) 
\int_{\bf R}
\phi^{4} ({\bf R}) .
\end{eqnarray}

To conclude this section, we summarize the results by remarking 
that the Wigner representation and the length scale considerations
bring the free energy in an almost-local form.  We need
to qualify that statement because of 
the appearance of the Laplacian, a non-local operator, in the 
$ h^{\phi} $ -contribution to the energy.
In fact, it is the non-locality of this term that gives
rise to a generalized Gross-Pitaevski or non-linear Schrodinger equation (NLSE).
The resulting (almost-local) ground state free energy is
$F = \int d^{3} R \; f({\bf R})$, where
\begin{eqnarray}
f({\bf R}) &=& \int_{\bf p} \; \left[ \frac{p^{2}}{2m} + V_{\rm{ext}} 
({\bf R}) - \mu \right] \rho({\bf R},{\bf p}) + v_{\rm{exch}} ({\bf R})
+ v_{\rm{dir}} ({\bf R}) + v_{\rm{pair}} ({\bf R}) \nonumber\\
& &+ \phi ({\bf R}) \left[ \frac{-\nabla ^{2}}{2m} + V_{\rm{ext}} 
({\bf R}) - \mu \right] \phi ({\bf R}) + 2 v_{\rm{exch}}^{\phi} ({\bf R})
+ 2 v_{\rm{dir}}^{\phi} ({\bf R}) - 2 v_{\rm{pair}}^{\phi} ({\bf R}) \nonumber\\
& & + \frac{1}{2} v(0) \phi^{4} ({\bf R}),
\end{eqnarray}
where the exchange, direct and pairing energy densities are the integrands
of the corresponding interaction energy contributions to the free energy :
\begin{eqnarray}
v_{\rm{exch}} ({\bf R}) &=& \frac{1}{2} 
\int_{\bf p} \; \int_{{\bf p}'} \; \rho ({\bf R},{\bf p})
v({\bf p}-{\bf p}') \rho_{W} ({\bf R},{\bf p}') \; , \nonumber\\
v_{\rm{pair}}({\bf R})  &=& \frac{1}{2} 
\int_{\bf p} \; \int_{{\bf p}'} \; \Delta_{W} ({\bf R},{\bf p})
v({\bf p}-{\bf p}') \Delta_{W} ({\bf R},{\bf p}') \; ,  \nonumber\\
v_{\rm{dir}} ({\bf R})  &=& \frac{1}{2} 
v(0) \int_{\bf p} \; \int_{{\bf p}'} \;
\rho_{W} ({\bf R},{\bf p}) \rho_{W} ({\bf R},{\bf p}') \; , \nonumber\\
v_{\rm{exch}}^{\phi} ({\bf R}) &=& \frac{1}{2} 
\phi^{2} ({\bf R})
\int_{\bf p}  \; \rho_{W} ({\bf R},{\bf p})
v({\bf p}) \; ,
\nonumber\\
v_{\rm{pair}}^{\phi}({\bf R})  &=& \frac{1}{2} 
\phi^{2} ({\bf R})
\int_{\bf p} \; \Delta_{W} ({\bf R},{\bf p})
v({\bf p}) \; ,
\nonumber\\
v_{\rm{dir}}^{\phi} ({\bf R})  &=& \frac{1}{2} 
\phi^{2} ({\bf R})
v(0) \int_{\bf p} \;
\rho_{W} ({\bf R},{\bf p}) \; .
\end{eqnarray}
Notice that the free energy and free energy density are functionals
of $\Delta ({\bf R},{\bf p})$, $\rho ({\bf R},{\bf p})$
and $\phi ({\bf R})$.  In the next section we determine 
the equilibrium values of  $\Delta ({\bf R},{\bf p})$,
$\rho ({\bf R},{\bf p})$ and  $\phi ({\bf R})$
by minimizing $F\left[ \rho, 
\Delta, \phi ; \mu \right]$.

\section{Self-Consistent Mean Field Theory}

In this section we derive the self-consistent mean-field 
equations that describe the
nearly-uniform bose condensate at zero temperature.  
In the variational method, one minimizes 
the mean-field ground state free 
energy $F\left[ \rho ,
\Delta , \phi ; \mu \right]$. 
Writing the integrands of the different contributions to the 
mean-field free energy in the 
Wigner representation, followed by the length scale arguments of the previous 
section showed that $F\left[ \rho_{W},
\Delta_{W}, \phi ; \mu \right]$ is essentially a local quantity.  Finally, in
the Thomas-Fermi limit of a nearly-homogeneous system,
$\rho_{W} ({\bf R},{\bf p})$ and $\Delta_{W} ({\bf R},{\bf p})$ 
are parametrized
by a single Bogolubov transformation parameter $\sigma ({\bf R},{\bf p})$
in the manner of Eq.(\ref{e:rds1}).  
Thus, to describe a nearly-homogeneous system, we minimize the Thomas-Fermi
ground state free energy, which is obtained from 
the mean-field free energy, 
assuming that $\rho_{W}$ and $\Delta_{W}$ are parametrized
by $\sigma$ (\ref{e:rds1}), 
$F \left[ \sigma, \phi ; \mu \right]$ $=$  $F\left[ \rho_{W}(\sigma),
\Delta_{W}(\sigma), \phi ; \mu \right]$.
We obtain the condensate wave function $\phi_{0} ({\bf R})$
and Bogolubov parameter $\sigma_{0} ({\bf R},{\bf p})$ that describe the
condensate 
by varying $\sigma$ and $\phi$ independently to get a minimum in F :
\begin{eqnarray}
\left. \frac{\delta F }{\delta \phi ({\bf R})} 
\right|_{\sigma = \sigma_{0},\phi=\phi_{0}} &=&
0 \; , \; \; \; \; (\rm{NLSE})\; \nonumber \\
\left. \frac{\delta F }
{\delta \sigma ({\bf R},{\bf p})} \right|_{\sigma = \sigma_{0},\phi=\phi_{0}}
&=&
0.
\label{e:sfe}
\end{eqnarray}
The $\phi$ variation, $\delta F / \delta \phi =0$, gives 
the non-linear Schrodinger Equation (NLSE).
The $\sigma$ variation, $\delta F / \delta \sigma = 0, $
gives an equation for $\sigma_{0} ({\bf R},{\bf p})$.
From $ \rho = \frac{1}{2} \left[ \cosh (2 \sigma) -1 \right]$ and
$\Delta = \frac{1}{2} \sinh (2 \sigma)$ (\ref{e:rds1}),
we find that
$\partial \rho / \partial \sigma = \sinh (2 \sigma)$
and $\partial \Delta / \partial \sigma = \cosh (2 \sigma)$,
so that  $\delta F / \delta \sigma = 0$ is equivalent to 
\begin{equation}
\tanh (2 {\sigma}_{0}) =
\frac{ - \delta F / \delta \Delta_{W}}
{\delta F / \delta \rho_{W}} .
\label{e:tan}
\end{equation}
Now, several terms of the NLSE, as well as the functional derivatives
$\delta F / \delta \Delta$ and $\delta F / \delta \rho$ (\ref{e:tan}), depend on
$\sigma_{0}$ and $\phi_{0}$ so that the resulting equations have to be solved
self-consistently.
To make the self-consistent nature of the equations more explicit, we consider
the $\sigma$-dependent contributions to the functional derivatives,
$\delta V_{\rm{exch}} / \delta \rho$,
$\delta V_{\rm{dir}} / \delta \rho$ and
$\delta V_{\rm{pair}} / \delta \Delta$, which we shall call
the generalized potentials,
\begin{eqnarray}
U_{\rm{exch}} ({\bf R},{\bf p}) &=&
\delta V_{\rm{exch}} / \delta \rho_{W}
({\bf R},{\bf p}) \; \; = 
\int_{{\bf p}'} \; v({\bf p}-{\bf p}') \rho_{W} ({\bf R},
{\bf p}') , \nonumber\\
\nonumber\\
U_{\rm{dir}} ({\bf R}) &=&
\delta V_{\rm{dir}} / \delta \rho_{W}
({\bf R},{\bf p}) \; \; = 
v(0) \int_{{\bf p}'} \; \rho_{W} ({\bf R},
{\bf p}') , \nonumber\\
\nonumber\\
U_{\rm{pair}} ({\bf R},{\bf p}) &=&
\delta V_{\rm{pair}} / \delta \Delta_{W}
({\bf R},{\bf p}) \; \; = 
\int_{{\bf p}'} \; v({\bf p}-{\bf p}') \Delta_{W} ({\bf R},
{\bf p}') ,
\label{e:genp}
\end{eqnarray}
where we name the generalized potentials after the respective
interaction
energies of which they are the functional derivatives,
$U_{\rm{exch}}$ is the exchange potential, $U_{\rm{dir}}$
the direct potential and $U_{\rm{pair}}$ the pairing potential.
Writing the distribution and pairing function in the integrands
of the generalized potentials in terms of $2\sigma$,
we find  with (\ref{e:tan}) that the generalized potentials 
implicitly depend on the functional derivatives of F :
\begin{eqnarray}
U_{\rm{exch}} ({\bf R},{\bf p}) &=&
\int_{{\bf p}'} \; v({\bf p}-{\bf p}')
\frac{1}{2} \left[
\frac{\delta F / \delta \rho}
{\sqrt{(\delta F / \delta \rho)^{2} - (\delta F / \delta \Delta)^{2}}}
-1 \right] , \nonumber\\
\nonumber\\
U_{\rm{dir}} ({\bf R}) &=&
v(0) \int_{{\bf p}'} \; \frac{1}{2} \left[
\frac{\delta F  / \delta \rho }
{\sqrt{(\delta F / \delta \rho)^{2} - (\delta F / \delta \Delta)^{2}}}
-1 \right] , \nonumber\\
\nonumber\\
U_{\rm{pair}} ({\bf R},{\bf p}) &=&
\int_{{\bf p}'} \; v({\bf p}-{\bf p}') \; \frac{1}{2} \left[
\frac{ - \delta F  / \delta \Delta}
{\sqrt{(\delta F / \delta \rho)^{2} - (\delta F / \delta \Delta)^{2}}}
\right], 
\label{e:genpo}
\end{eqnarray}
where it is understood that the functional derivatives in the integrands
are evaluated at ${\bf R}$ and ${\bf p}'$.
Functional differentiation shows that the functional derivatives of F
in turn depend on the generalized potentials,
\begin{eqnarray}
\frac{\delta F}{\delta \Delta_{W} ({\bf R},{\bf p})} &=&
U_{\rm{pair}} ({\bf R},{\bf p}) - \phi^{2} ({\bf R})
v({\bf p}) \; , \nonumber\\
\nonumber\\
\frac{\delta F}{\delta \rho_{W} ({\bf R},{\bf p})} &=&
\frac{p^{2}}{2m} + V_{\rm{ext}} ({\bf R}) - \mu \nonumber\\
&& + U_{\rm{exch}} ({\bf R},{\bf p}) + U_{\rm{dir}}
({\bf R},{\bf p}) + \phi^{2} ({\bf R}) \left[ v({\bf p})+v(0)
\right] \; . 
\label{e:fund}
\end{eqnarray}
Thus, equations (\ref{e:genpo}) and (\ref{e:fund})
self-consistently determine the generalized potentials.
Furthermore, there is a dependence on the condensate wave function $\phi$.
The latter has to be obtained from the NLSE :
\begin{equation}
\left[ - \frac{\nabla^{2}}{2m} + V_{\rm{ext}} ({\bf R})
- \mu + U({\bf R}) + v(0) \phi^{2} ({\bf R}) \right] \phi({\bf R}) = 0 , 
\label{e:nlse}
\end{equation}
where the potential $U({\bf R})$, is equal to :
\begin{equation}
U({\bf R}) =  U_{\rm{dir}} ({\bf R}) + U_{\rm{exch}} ({\bf R},0)
- U_{\rm{pair}} ({\bf R},0) \; .
\label{e:unlse}
\end{equation}
This potential term, which stems from the interaction of the condensate
with the particles out of the condensate, is absent in the simplest
(low density limit) form of the NLSE, usually encountered in the 
literature.

The equations,  (\ref{e:genpo}) (\ref{e:fund}) (\ref{e:nlse}) and 
(\ref{e:unlse}), are the full set of self-consistent mean field equations
that describe the condensate in the Thomas-Fermi approximation.
The self-consistent equations for the homogeneous gas 
\cite{GR}, are recovered by 
putting $V_{\rm{ext}} = 0$ and by assuming that $\phi$ is independent
of position so that the kinetic energy contribution to the NLSE vanishes.
Regarding the connection with the intuitive Thomas-Fermi model, we note
that $\mu$ and $V_{\rm{ext}}$ in the self-consistent mean field equations
always appear as $\mu - V_{\rm{ext}} ({\bf R})$, so that it is natural
to define a local effective chemical potential:
\begin{equation}
\mu_{\rm{eff}} ({\bf R}) = \mu - V_{\rm{ext}} ({\bf R}).
\label{e:mueff}
\end{equation}
In fact, this is the essence of the Thomas-Fermi description : the system
is described locally as a homogeneous system with a position dependent
effective chemical
potential (\ref{e:mueff}).
  
The solutions to the fully self-consistent equations
determine the expectation
value of all (static) physical observables as a function of the 
chemical potential $\mu$.  One observable we can obtain 
in this manner is N, the number of trapped particles,
\begin{equation}
N(\mu) = \frac{\partial F}{\partial \mu} = \int_{\bf R} \int_{\bf p} 
\rho ({\bf R},{\bf p})
+ \int_{\bf R} \; \phi^{2} ({\bf R}) ,
\end{equation}
the inversion of which yields $\mu$(N), from which we can
cast the results for the thermodynamic quantities in terms of the 
parameter that is controled
or measured in the experiment $-$ the number of atoms N.

\section{Low Density Limit}

The self-consistent equations,
(\ref{e:genpo}) (\ref{e:fund}) (\ref{e:nlse}) and
(\ref{e:unlse}) can be solved iteratively.
In the low density regime, where $\sqrt{n a^{3}} \ll 1$, we 
approximate the result by the expressions obtained after 
a single iteration,
starting from  $\sigma^{(0)}_{0} =0$ ($U^{(0)}_{\rm{exch}}= $
$U^{(0)}_{\rm{dir}} = $ $U^{(0)}_{\rm{pair}} = 0$, where
the superscript indicates the order of the iteration).
With this first guess we solve the NLSE and obtain the
functional derivatives (\ref{e:genpo}), $\delta F/\delta \rho$,
$\delta F/ \delta \Delta$, yielding
the first-order $\sigma -$parameter (\ref{e:tan}), $\sigma^{(1)}$,
and the generalized potentials (\ref{e:genp}), $U^{(1)}_{\rm{dir}}$,
$U^{(1)}_{\rm{exch}}$, $U^{(1)}_{\rm{pair}}$.  With these
single iteration expressions we compute the expectation values of
the observable quantities.

In solving the NLSE, we shall assume that $\phi ({\bf R})$ varies slowly enough
that we can also neglect the kinetic energy operator. 
To make the dependence
on the scattering length explicit, we replace the potential
by a pseudopotential,
\begin{equation}
V_{\rm{pseudo}} ({\bf r}) = \lambda \delta({\bf r}) \frac{\partial}
{\partial r} r ,
\label{e:pseudo}
\end{equation}
where $\lambda = 4 \pi \hbar^{2} a /m$ and the derivative operator 
is necessary to remove the divergency in the ground state free energy
\cite{HY}.

Furthermore, we shall assume that $\phi ({\bf R})$ varies slowly enough
that we can also neglect the kinetic energy operator in solving the NLSE
(\ref{e:nlse}) :
\begin{equation}
\lambda \left[ \phi^{{(1)}} ({\bf R}) \right] ^{2}
= \mu_{\rm{eff}} ({\bf R}) , 
\label{e:phio}
\end{equation}
where $\mu_{\rm{eff}}$ is the effective chemical potential 
(\ref{e:mueff}).
The functional derivatives (\ref{e:fund}) are 
\begin{eqnarray}
\frac{\delta F^{\rm{{(1)}}}}{\delta \Delta} &=&
- \lambda \left[ \phi^{{(1)}} ({\bf R}) \right] ^{2} \; , \nonumber\\
\nonumber\\
\frac{\delta F^{\rm{{(1)}}}}{\delta \rho} &=&
\frac{p^{2}}{2m} - \mu_{\rm{eff}} ({\bf R}) + 
2 \lambda  \left[ \phi^{{(1)}} ({\bf R}) \right] ^{2}. 
\end{eqnarray}
Consequently, the single iteration value for the Bogolubov
transformation parameter $\sigma$ is equal to 
\begin{eqnarray}
\tanh \left[ 2 \sigma^{\rm{{(1)}}}_{0} ({\bf R},{\bf p}) \right]
&=& \frac{ \lambda \phi^{2} ({\bf R})}
{(p^2/2m) - \mu_{\rm{eff}} ({\bf R}) + 
2 \lambda  \phi^{2} ({\bf R})} \nonumber\\
\nonumber\\
&=& \frac{\mu_{\rm{eff}} ({\bf R})}
{(p^2/2m) + \mu_{\rm{eff}} ({\bf R})} \; \; ,
\label{e:sigm1}
\end{eqnarray}
which can be recognized as the dilute uniform gas result if we put
$\mu_{\rm{eff}} = \mu$.

The expression for the Bogolubov parameter $\sigma^{{(1)}}_{0}$
from Eq.(\ref{e:sigm1}) is what we would have obtained with 
an effective energy density neglecting the interaction energies of the
particles out of the condensate,
$V_{\rm{dir}} $, $ V_{\rm{exch}} $ and
$ V_{\rm{pair}} $.  In other words, the effective ground state
energy is
$F_{\rm{eff}}$ = $\int d^{3} R \; f_{\rm{eff}} ({\bf R})$, where
\begin{eqnarray}
f_{\rm{eff}} ({\bf R}) &=& 
\int_{\bf p}  \; \left[ 
\left[ \frac{p^{2}}{2m} - \mu_{\rm{eff}} ({\bf R}) 
+ 2 \lambda \phi^{2} ({\bf R}) \right] \;
\rho_{W} ({\bf R},{\bf p})
- \lambda \phi^{2} ({\bf R}) \Delta_{W} ({\bf R},{\bf p}) \right]
\nonumber\\
&& - \mu_{\rm{eff}} ({\bf R})  \phi^{2} ({\bf R})
+ \frac{\lambda}{2} \phi^{4} ({\bf R}) .
\label{e:feff}
\end{eqnarray}
We obtain the results for the observable quantities
by calculating their expectation values 
from the single iteration $\sigma^{{(1)}}_{0}$ of Eq.(\ref{e:sigm1}).
For example, the condensate wave function is determined from the
NLSE :
\begin{equation}
\lambda \phi^{2} ({\bf R}) \approx
\mu_{\rm{eff}} ({\bf R}) - U^{(1)} ({\bf R}), 
\label{e:gap1}
\end{equation}
where the potential $U({\bf R})$ is the sum of the generalized potentials
at zero momentum (\ref{e:unlse}),
\begin{equation}
U^{(1)} ({\bf R}) =  U^{(1)}_{\rm{exch}}
({\bf R},0) +  U^{(1)}_{\rm{dir}} ({\bf R})
- U^{(1)}_{\rm{pair}} ({\bf R},0) ,
\nonumber\\
\end{equation}
evaluated with the single-iteration value for $\sigma$.
The single-iteration values for the generalized potentials are computed to be :
\begin{eqnarray}
U^{(1)}_{\rm{exch}} ({\bf R},0) &=&
U^{(1)}_{\rm{dir}} ({\bf R}) = 
\frac{\lambda}{3\pi^2} \left[ \mu_{\rm{eff}} ({\bf R}) \right]^{3/2} 
m^{3/2} \; , \nonumber\\
U^{(1)}_{\rm{pair}} ({\bf R},0) &=&
-{\lambda\over\pi^2} \; \left[ \mu_{\rm{eff}} ({\bf R}) \right]
^{3/2} m^{3/2} \; . 
\label{e:u1}
\end{eqnarray}
Thus, the condensate density is (\ref{e:gap1}) 
\begin{equation}
\phi^{2} ({\bf R}) \approx {1\over\lambda}\mu_{\rm eff} ({\bf R})
- \frac{5}{3 \pi^{2}} \left[\mu_{\rm{eff}} ({\bf R}) \right] ^{3/2}
m^{3/2}\; \; .
\label{e:phi1}
\end{equation}
The total density n(${\bf R}$), including the correction
to $\phi^{2} ({\bf R})$ (\ref{e:phi1}) and the local depletion, is 
equal to
\begin{eqnarray}
n({\bf R}) &=&
\phi ^{2} ({\bf R}) + \; \int_{\bf p} \;
\rho ({\bf R},{\bf p})\nonumber\\
&\approx& \phi ^{2} ({\bf R}) + \frac{1}{3\pi^2}
\left[\mu_{\rm{eff}} ({\bf R}) \right] ^{3/2}
m^{3/2}\nonumber\\
&\approx& {1\over\lambda}\mu_{\rm{eff}} ({\bf R})
- \frac{4}{3 \pi^{2}} \left[\mu_{\rm{eff}} ({\bf R}) \right] ^{3/2}
m^{3/2} ,
\label{e:npert}
\end{eqnarray}
resulting in an expression for the density, n(${\bf R}$),
in terms of the effective
chemical potential $\mu_{\rm{eff}} ({\bf R})$.  Inverting this
relation up to first order in $\sqrt{n a^{3}}$, we obtain 
\begin{equation}
\mu_{\rm{eff}} ({\bf R}) \approx \lambda n({\bf R})
\left[ 1 + \frac{32}{3} \sqrt{\frac{n({\bf R}) a^{3}}{\pi}} 
\right] ,
\label{e:mu1}
\end{equation}
which, for the homogeneous case, reduces to the well-known perturbation result. 

Finally, in a similar manner, we obtain the local pressure
$P({\bf R})$ from
the expression for the effective free energy density (\ref{e:feff}),
$P({\bf R})$ = $ - f^{(1)}_{\rm{eff}} ({\bf R})$,
\begin{equation}
P({\bf R}) = \frac{\lambda \phi^{4} ({\bf R})}{2}
\left[ 1 - \frac{128}{15\pi^2} \sqrt{ n({\bf R}) a^{3}} \right]. 
\label{e:pres}
\end{equation}
We can then replace $\phi^{2}$ in (\ref{e:pres}) by its 
single-iteration value (\ref{e:phi1}). Furthermore, replacing
$\mu_{\rm{eff}} ({\bf R})$ in the resulting expression 
by (\ref{e:mu1}) results in a local equation of state.

The above results illustrate an important advantage of the Thomas-Fermi
description $-$ by neglecting the kinetic energy operator in the NLSE we
recover simple analytical expressions
for most quantities.  These expressions are the analogues of the 
perturbation results for the dilute homogeneous bose gas.
It is then of course very important to determine
the regime and the conditions under which these results can be trusted.

One source of error in the theory stems from neglecting the Laplacian operator
in the NLSE.  
This approximation, although convenient, is
not part of the Thomas-Fermi description. 
It is always
possible to calculate the condensate wave function numerically from the NLSE 
and proceed from there with the iteration of the 
self-consistent Thomas-Fermi
equations (\ref{e:sfe})! 
Nevertheless, if we omit 
the Laplacian term, we can estimate the error by calculating $e_{L}$, the ratio
of the kinetic energy term, $-\nabla^{2} \phi /2m$, and the non-linear
potential energy in the NLSE, $\lambda \phi^{3}$,
 \begin{equation}
e_{L} ({\bf R}) = | -\nabla^{2} \phi /  2m \lambda \phi^{3} | = 
\left| \frac{ -\nabla^{2} \phi ({\bf R}) / \phi 
({\bf R}) }{k_{c}^{2}({\bf R})} \right|,
\label{e:el}
\end{equation}
where $k_{c}({\bf R})$ = $\left[ 8 \pi a n({\bf R}) \right] ^{\frac{1}{2}}$ is
the inverse of the local coherence length, $k_{c}$
= $\lambda_{c}^{-1}({\bf R})$.
 
Another source of error, which cannot be remedied but is truly inherent to the
Thomas-Fermi approximation, stems from the inhomogeneity of the system.
This error is also more difficult to estimate, and one benefit of our
approach is that the gradient expansion offers a `handle' on this quantity.
Indeed, we use the lowest order non-vanishing term in the gradient expansion
to estimate the error.  This term is of second order because the first order
term vanishes. 
We estimate its magnitude (\ref{e:gradexp}) by 
replacing the partial derivatives with respect to the momentum
variables by $k_{c}^{-1}$, since $k_{c}$ is a measure of the range
in ${\bf p}$ of the observable at zero temperature. The 
relative error for the general product of two arbitrary operators
A and B, $e_{\rm{i}} \left[ AB \right] $, is then given by 
\begin{equation}
e_{\rm{i}} \left[ AB \right] \approx 
\frac{1}{8 k^{2}_{c} ({\bf R})}
\left[ \frac{\nabla^{2} A_{W} }{ A_{W} } + \frac{ \nabla^{2} B_{W} }{ B_{W}} 
- 2 
\frac{ \nabla A_{W}\; \cdot \nabla B_{W} }{ A_{W} B_{W} } \right] ,
\end{equation}
The validity of the Thomas-Fermi description depends on 
$X_{W}^{2}-Y_{W}^{2}=1$
(\ref{e:canw}), so that we use the accuracy of this equality to test the
validity of the local homogeneity description. 
The expression (\ref{e:canw}) can also be written as
$\exp \left[ \sigma({\bf R},{\bf p}) \right] 
\exp \left[ -\sigma({\bf R},{\bf p}) \right] =1 $, so that we choose
$A_{W}$ as $\exp \left[ \sigma({\bf R},{\bf p}) \right]$ and
$B_{W}$ as $\exp \left[ -\sigma({\bf R},{\bf p}) \right]$ to estimate
the relative error $e_{\rm i}$.  In fact, it is more
convenient to work with $\exp(4\sigma)$ then $\exp(\sigma)$, so that we
compute the inhomogeneity error $e_{i} \left[ 
\exp(4\sigma) \exp(-4\sigma)\right] $ of $\exp(4\sigma)$ and divide by 4 
(since the relative error of $f^{n}$ is simply n $\times$ the relative error
of f).  
In this manner, we find that
\begin{eqnarray}
e_{i} \left[ \exp(\sigma) \exp(-\sigma) \right] &=&
\frac{1}{4} e_{i} \left[ \exp(4\sigma) \exp(-4\sigma) \right]
\nonumber\\
&\approx& \frac{1}{8 k_{c}^{2} ({\bf R})} \left| \frac{\nabla \exp(4\sigma)}
{\exp(4\sigma)} \right| ^{2} \; .
\label{e:inh}
\end{eqnarray}
With the single-iteration value for the low-density condensate,
\begin{equation}
\exp \left[ 4\sigma({\bf R},{\bf p}) \right] = 
1 + \frac{2 \mu_{\rm{eff}} ({\bf R})}{p^{2}/2m} ,
\end{equation}
we find that the inhomogeneity error, $e_{i} ({\bf R})$ (\ref{e:inh}),
is equal to
\begin{equation}
e_{i} ({\bf R}) = \frac{1}{2} \left| 
\frac{{\bf F}_{\rm{ext}}({\bf R}) \lambda_{c}
({\bf R})} {(p^{2}/{2m}) + 2 \mu_{\rm{eff}} ({\bf R})} \right| ^{2} ,
\end{equation}
where ${\bf F}_{\rm{ext}}$ is the force of the external potential,
 ${\bf F}_{\rm{ext}}$ = $- \nabla V_{ext}$.
As expected, the error is largest for ${\bf p} = 0$, and using the 
 ${\bf p} = 0$ $-$ value, we obtain a simple position dependent estimate
for the inhomogeneity error, $e_{i} ({\bf R})$,
\begin{equation}
e_{i} ({\bf R}) = \frac{1}{8} |{\bf F}_{\rm{ext}}({\bf R})
\lambda_{c}({\bf R}) / \lambda \phi^{2} ({\bf R}) |^{2} ,
\label{e:inhf}
\end{equation} 
where we replaced $\mu_{\rm{eff}}$ by $ \lambda \phi^{2}$.
By equating this error (\ref{e:inhf}) to a chosen value, $e_{\rm{cut}} \ll 1$,
reflecting the accuracy we demand from the theory, we can
determine the spatial boundary beyond which the Thomas Fermi Theory 
is less accurate than $e_{\rm{cut}}$.

\section{Spherically Symmetric Harmonic Oscillator Trap}

We now specialize $V_{\rm{ext}} ({\bf R})$ to a harmonic oscillator potential,
\begin{equation}
V_{\rm ext} ({\bf R}) = \frac{1}{2} \hbar \omega (R/L)^2 \; ,
\end{equation}
where $L$ is the size of the harmonic oscilator ground state,
\begin{equation}
L = \sqrt{\frac{\hbar}{m \omega}},
\end{equation}
and compute the expectation value of important quantities
in the low density limit of the previous section.

In zeroth order in the iteration, we recover the results of Baym and Pethick
\cite{GB}.  From (\ref{e:phio}) we see that
\begin{eqnarray}
\left[ \phi^{(0)} ({\bf R}) \right] ^{2} 
&=& \left[ \mu - V_{\rm{ext}} ({\bf R}) \right] / \lambda 
\nonumber\\
&=& \frac{R^{2}_{0}}{8 \pi a L^{4}} \left[ 1 - (R/R_{0})^{2} \right] 
\label{e:phios}
\end{eqnarray}
where $R_{0}$ is the size of the condensate, 
$R_{0} = \sqrt{\frac{2 \mu}{\hbar \omega}}
L$.  In zeroth order, all particles are in the condensate, so that 
N $= \int_{\bf R} \phi^{2} ({\bf R}) $, and 
\begin{equation}
\mu^{(0)} = \frac{\hbar \omega}{2} \left(\frac{ 15 a N}{L} \right) ^{2/5} ,
\end{equation} 
and consequently,
\begin{equation}
R_{0} = L \left( \frac{15 a N}{L} \right) ^{1/5} .
\end{equation}
The local coherence length, $\lambda_{c} ({\bf R})$ is given by
\begin{equation}
\lambda_{c} ({\bf R}) = \frac{L^{2}}{\sqrt{R_{0}^{2} - R^{2}}} \; \; .
\label{e:cohl}
\end{equation}

Before we proceed to calculate the perturbation corrections to the
observables, we test the validity of the low density Thomas-Fermi
formalism by calculating the errors.  The error due to neglecting the
Laplacian in the NLSE, $e_{L} ({\bf R})$ (\ref{e:el}),
is easily computed with
(\ref{e:cohl}):
\begin{equation}
e_{L} ({\bf R}) = \left( \frac{L}{R_{0}} \right) ^{4} 
\frac{ \left[ 3  - 2 (R/R_{0})^{2} \right] }{(1 - (R/R_{0})^{2})^{3}} \; \; ,
\end{equation}
from which we see that the laplacian can be omitted in the NLSE on
condition that the size of the condensate is much larger than the size
of the ground state, $ R_{0} \gg L $, or $ \left( 15 a N / L \right)^{1/5} 
\gg 1$.  
The error due to the departure of the BEC from homogeneity, $e_{i} ({\bf R})$
(\ref{e:inh}), is
\begin{equation}
e_{i} ({\bf R}) = \frac{1}{2} \left( \frac{L}{R_{0}} \right) ^{4} 
\frac{(R/R_{0})^{2}}{(1 - (R/R_{0})^{2})^{3}} \; \; . 
\end{equation}
Again, notice that $e_{i}$ is small over most of the condensate region
($R < R_{0}$) if  $ R_{0} \gg L $.

In Fig.~(1) we show the density, $[\phi^{(0)} (R)]^{2} 
/ [ \phi^{(0)} (R=0) ]^{2}$ (\ref{e:phios}), and both errors,
$e_{L}$ and $e_{i}$, as a function
of the distance to the middle of the trap. 
\begin{figure}[htbp]
\centerline{\BoxedEPSF{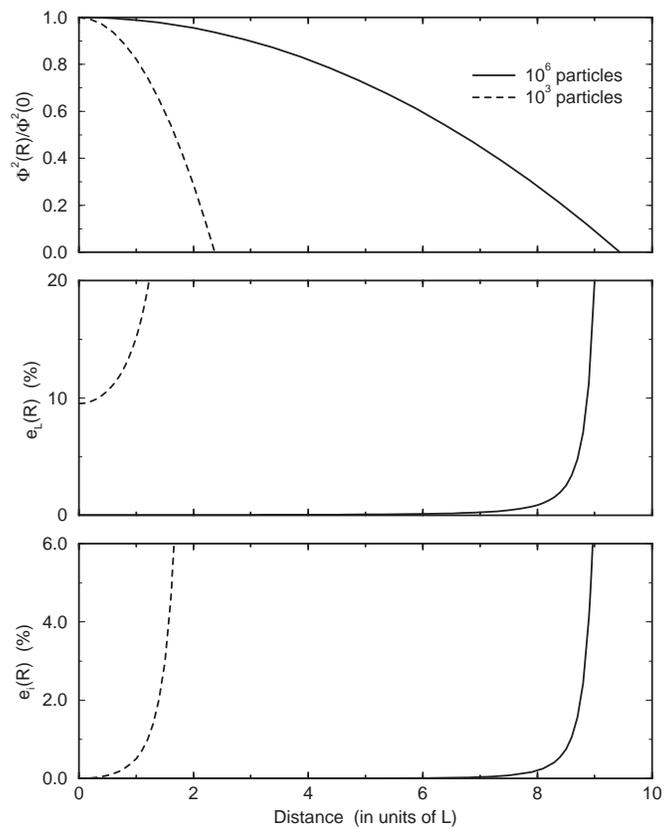 scaled 600}}
\caption{{\sf (a) Condensate density for N = $10^{3}$ and $10^{6}$; (b) Error
incurred in neglecting kinetic term in NLSE; (c) Error incurred in
Thomas-Fermi approximation.  Length scale on horizontal axis is in units of L,
the extend of the ground state wave function.  Calculations are done for
L = $10^{-4}$ cm, scattering length a = $5 \times 10^{-7}$ cm.}}
\end{figure}
The curves are
calculated for a harmonic oscillator trap
of $L = 1 \mu m$ and an inter-atomic interaction with
scattering length a = 5 nm.  The dotted lines
correspond to $N = 10^{3}$ atoms in the trap, and the full line gives the
results for $N = 10^{6}$ atoms.  Notice that for $10^{3}$ particles, 
the Laplacian error is already substantial ($\sim 10 \%$) in the middle of
the trap.  In contrast, only at 
$ R = 1.8 L $ (to be compared to $ R_{0} = 2.4 L $) does the
inhomogeneity error become of comparable magnitude.  This indicates
that even for as few as 1000 particles in the trap, the Thomas-Fermi
description could be reasonably accurate for these parameters, 
provided one keeps 
the kinetic energy term
in solving the NLSE.  For $10^{6}$ atoms, $e_{i}$ and $e_{L}$ only become of
the order of $ 10 \%$ at $ R = 9.0 L$, whereas $ R_{0} = 9.4 L$, which
shows 
that the Thomas-Fermi description and neglecting the Laplacian operator
are valid approximations in almost all of the condensate region. 

Under this condition, it is meaningful to calculate the perturbation 
corrections to the expectation values of the observable quantities.
Including the perturbation correction, the local density (\ref{e:npert})
is equal to 
\begin{equation}
n(R) = \frac{R_{0}^{2}}{8 \pi a L^{4}}
\left[ 1 - (R/R_{0})^{2} \right] 
\left[ 1 - \frac{2 \sqrt{2}}{3 \pi} \frac{a R_{0}}{L^{2}} 
\sqrt{1 - 
\left(R/R_{0}\right)^{2} } \right].
\end{equation}
The number of trapped particles, N, is obtained by integrating over the
density $n({\bf R})$,
\begin{equation}
N = \int_{\bf R} \; n(R) = 4 \pi \int_{0}^{R_{0}} d R \; R^{2} \; n(R) \; ,
\label{e:nint}
\end{equation}
which leads to
\begin{equation}
N = \frac{1}{15} \frac{L}{a} \left(\frac{2 \mu}{\hbar \omega}\right)^{5/2}
- \frac{\sqrt{2}}{24}  \left(\frac{2 \mu}{\hbar \omega}\right)^{3} .
\end{equation}
The inverse relation, $\mu$ as a function of N, can be obtained
by solving for $\mu$ iteratively in the previous equation,
wich gives up to second iteration, the following result :
\begin{equation}
\mu = \frac{\hbar\omega}{2} \left(\frac{15 a}{L} \right)^{2/5} N^{2/5} \left[
1 + \frac{\sqrt{2}}{60}   \left(\frac{15 a}{L} \right)^{6/5} N^{1/5} \right].
\end{equation}
Similarly, we obtain the condensate density from (\ref{e:phi1}) or the local 
depletion, $d({\bf R}) = \left[ n^{(1)} ({\bf R}) \right.$
$-\left.  
\left[ \phi^{1} ({\bf R}) \right] ^{2} \right]
/ \left[ \phi^{(1)} ({\bf R}) \right] ^{2} $ :
\begin{equation}
d(R) = \frac{2 \sqrt{2}}{3 \pi} \frac{a R_{0}}{L^{2}}
\sqrt{ 1 - ( R/R_{0} )^{2} } .
\end{equation}
In Fig.~(2), we show the local depletion as a function of position
for the same parameters as those of Fig.~(1).  
\medskip
\begin{figure}[htbp]
\centerline{\BoxedEPSF{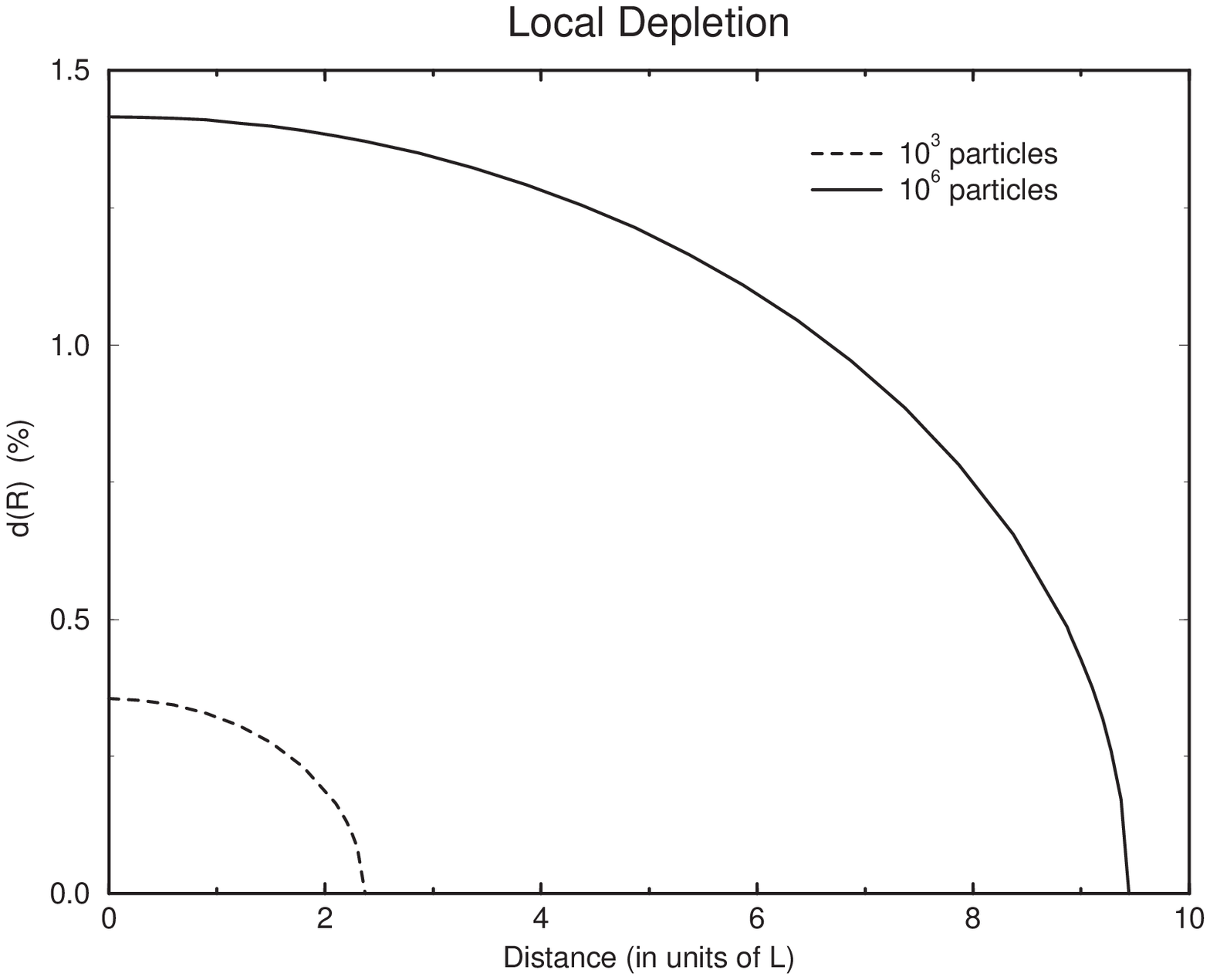 scaled 600}}
\caption{{\sf Depletion, defined as $d(R) = \left[ n({\bf R}) - \phi^{2} ({\bf R})
\right] / \phi^{2} ({\bf R})$, for the same systems as Fig.~(1).}}
\end{figure}
\medskip
The local pressure is
shown in Fig.~(3).
\medskip
\begin{figure}[htbp]
\centerline{\BoxedEPSF{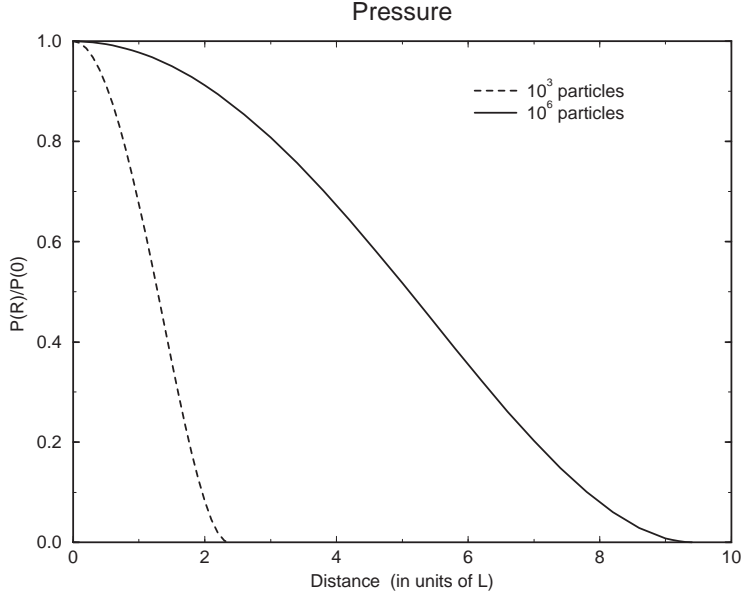 scaled 600}}
\caption{{\sf Pressure for the same systems as Fig.~(1).}}
\end{figure}
\medskip

To conclude this section, we repeat that the condition for
the validity of the Thomas-Fermi description is that the size
of the condensate exceeds the size of the ground state of the trap,
$R_{0} \gg L$.  An equivalent condition is that the coherence length
in the middle of the condensate is smaller than the size of the ground
state $\lambda_{c} (R=0) \ll L$, or that the chemical potential exceeds the
ground state energy, $\mu \gg (\hbar \omega /2)$.  These statements do not
depend on the details of the trapping potential. Of course, the shape of the
condensate, the boundary where the Thomas-Fermi description breaks down,
and the expectation values of the
local observables do depend on the shape of the potential.
In this section, we gave the results for a spherically symmetric harmonic
oscillator potential. 
For the convenience of the reader we tabulate several
of the results up to first non-vanishing order in table I.

\section{Density of States}

In the Thomas-Fermi picture, the system is locally equivalent to a 
uniform system.  Therefore, there are `local' excitations 
which in the low-density regime are described by the following
energy spectrum :
\begin{equation}
{\epsilon}_{p} ({\bf R}) =
\sqrt{(p^{2}/2m+\mu_{\rm{eff}}({\bf R}))^{2}-\mu_{\rm{eff}}^{2} ({\bf R}) }
+ \mu \; ,
\label{e:bspec}
\end{equation}
which is well known from the Bogolubov treatment of the uniform case.
The local dispersion relation (\ref{e:bspec}) describes a phonon with
position dependent sound velocity.

To obtain the excitation of the whole system we compute the density of states
using the formula
\begin{equation}
g(\epsilon) = \sum_{i} \; \delta(\epsilon - \epsilon_{i}),
\label{e:dsc}
\end{equation}
where $\sum_{i}$ represents the sum over all excited states.
In the spirit of the Thomas-Fermi approximation we take 
\begin{equation}
g(\epsilon) = \int_{\bf R} \int_{\bf p} \;
\delta \left(\epsilon - \epsilon_{\bf p} ({\bf R}) \right) \; .
\label{e:dstf}
\end{equation}
After integration over the momentum variable, we obtain 
\begin{equation}
g(\epsilon) = \frac{1}{2 \pi^{2}} \int_{\bf R} \; p_{\epsilon}^{2}
({\bf R}) \; \left| \frac{\partial \epsilon}{\partial p} \right| ^{-1} ,
\label{e:dss}
\end{equation}
where $p_{\epsilon} ({\bf R})$ is the momentum of a particle at position
${\bf R}$ with energy $\epsilon$.
When calculating the remaining integral over space, we need to distinguish
between spatial region (I) with condensate and a second region (II) without
condensate, shown schematically in Fig.~(4).  
\medskip
\begin{figure}[htbp]
\centerline{\BoxedEPSF{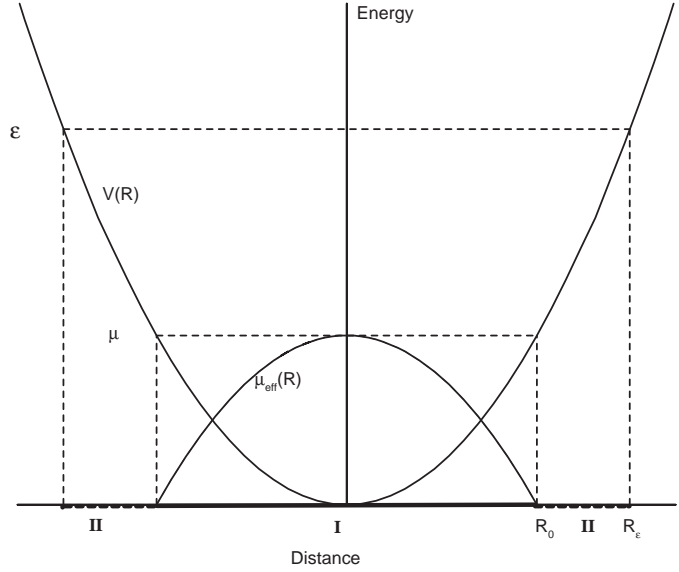 scaled 600}}
\caption{{\sf Schematic representation of the region with (region I), 
and without condensate (region II)
for a BEC in a harmonic trap.  The condensate density is
proportional to $\mu_{\rm{eff}} ({\bf R})$, which is a `mirror image' of
the trapping potential.  Particles in the condensate have energy $\mu$ and
a particle excited up to energy $\epsilon$ can move into region (II) as far
as the classical turning point $R_{\epsilon}$.}}
\end{figure}
\medskip
It is necessary to break
up the integral (\ref{e:dss}) over the different integration regions, 
because the dispersion relations for the excitations are different. 
In region (I), we use the Bogoliubov spectrum
(\ref{e:bspec}), whereas in region (II), the atoms are essentially free 
particles moving in the trap:
\begin{equation}
\epsilon_{p} ({\bf R}) = \frac{p^{2}}{2m} + V_{\rm{ext}} ({\bf R}) \; .
\label{e:fspec}
\end{equation}
The density of states is then the sum of the integrals over region (I) 
and (II):
\begin{eqnarray}
g(\epsilon) = \frac{\sqrt{2}}{2} \frac{m^{3/2}}{\pi^{2}} 
& & \left[
(\epsilon - \mu) \int_{(I)} d^{3}R \;
\frac{\sqrt{ \sqrt{\left[ \epsilon - \mu \right] ^{2} + 
\mu^{2}_{\rm{eff}}({\bf R})} - \mu_{\rm{eff}} ({\bf R})}} 
{\sqrt{\left[ \epsilon - \mu \right] ^{2} +
\mu^{2}_{\rm{eff}}({\bf R})}}
\right.
\nonumber\\
&& 
\left. 
+ \int_{(II)} d^{3} R \; \sqrt{\epsilon
- V_{\rm ext}({\bf R})} \right] .
\end{eqnarray}

For the special case of a spherically symmetric harmonic oscillator
trap, we find the following expression for the density of states :
\begin{eqnarray}
g(\epsilon) = \frac{4}{\pi} \; \frac{\mu^{2}}{(\hbar \omega)^{3}}
& & \left[ (\epsilon / \mu -1) \int_{0}^{1} dr \sqrt{1 - r}
\frac{ \sqrt{ \sqrt{ (\epsilon / \mu -1)^{2} + r^{2}} - r }}
{ \sqrt{ (\epsilon / \mu -1)^{2} + r^{2}}}
\right.
\nonumber\\
& & \left. \; + \; 2 \int_{1}^{\epsilon / \mu} dr \; r^{2} \; 
\sqrt{ \epsilon / \mu - r^{2}} \; 
\right] \; .
\end{eqnarray}
In Figs. (5) and (6) we show the density of states for the system discussed
in the previous section, L = $ 1 \mu m$, a = 5 nm, N = $10^{3}$ (Fig.~(5)) and
N = $10^{6}$ (Fig.~ (6)). 
\medskip
\begin{figure}[htbp]
\centerline{\BoxedEPSF{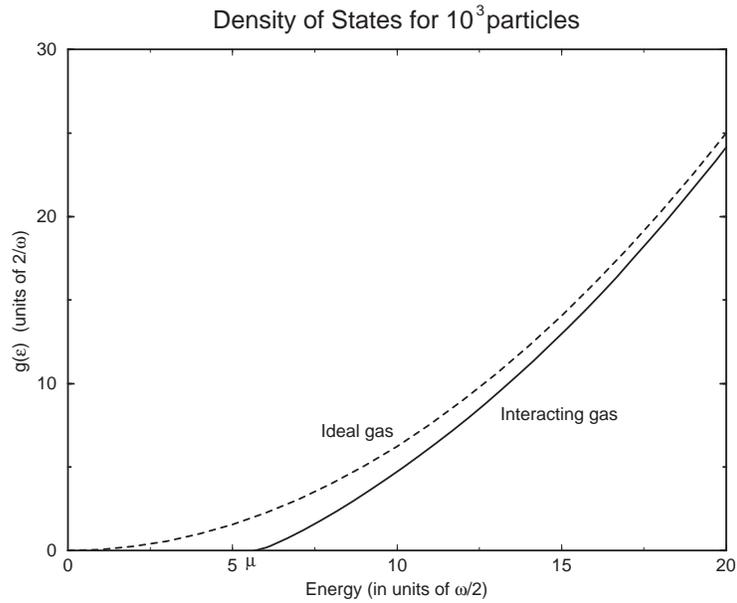 scaled 600}}
\caption{{\sf Density of states calculated in the Thomas-Fermi approach described
in the paper.  The system is a BEC of N = $10^{3}$ particles interacting
with a scattering length a = $5 \times 10^{-7}$ cm, 
in a harmonic trap with ground state of extend L = $10^{-4}$ cm.}}
\end{figure}
\medskip
\medskip
\begin{figure}[htbp]
\centerline{\BoxedEPSF{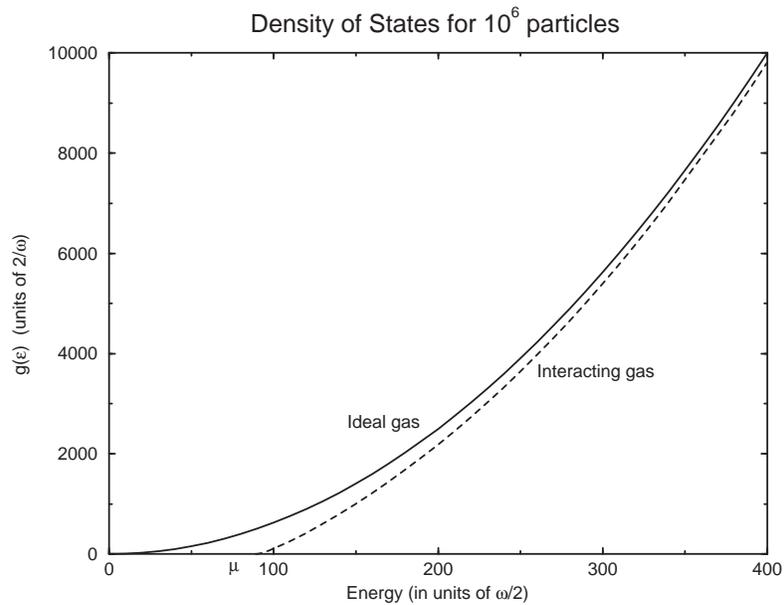 scaled 600}}
\caption{{\sf Density of states for the same system as in Fig.~(6), but with N = 
$10^{6}$ particles.}}
\end{figure}
\medskip
The dotted lines show the result for 
the interacting bose gas, the full line shows the density of states of
the ideal gas in the same trap. 
The density of states starts from the chemical potential
$\mu$, consistent with (\ref{e:bspec}), which implies that the energies
are measured from the bottom of the potential well so that a particle
of zero momentum in the condensate has energy $\mu$.  If we were to set out the
density of states as a function of excitation energy $\epsilon - \mu$,
the density of states curves for the interacting BEC-systems
would be shifted to the left by an amount $\mu$.
In contrast to the homogeneous BEC, the density of states
for the interacting case, as a function of the excitation energy, grows
faster than the density of states of the ideal gas.  The reason 
is purely geometrical: 
the phonon 
has a much larger volume in coordinate space available
(at least the volume of the condensate) 
than the non-interacting boson that received the same amount of energy
and can only move near the bottom of the potential well. 
This effect outweighs the fact that
the momentum space volume available to the phonon 
is less than the momentum space volume available to the non-interacting
particle with the same energy.  

Of course, the sharpness of the boundary between region (I)
and (II),
is an artifact of neglecting the Laplacian operator in the NLSE. 
Nevertheless, except for a region near the boundary, we argue
that the rest of space is well-described and that 
the contribution of the near-boundary region is comparatively
small so that the error that is introduced in the integral (\ref{e:dss})
is small provided the Thomas-Fermi
description is valid in most of the condensate region.

\section*{Acknowledgments}

This work was supported in part by funds provided by 
the U.S. Department of Energy under cooperative agreement 
\# DE-FC02-94ER40818. P.T. was supported by Conselho Nacional de 
Desenvolvimento Cientifico e Tecnologico (CNPq), Brazil.
The work of E.T. is supported by the NSF through a grant for the
Institute for Atomic and Molecular Physics at Harvard University
and Smithsonian Astrophysical Observatory.

\newpage

\centerline{\Large Table I}
\vskip 0.10in
\centerline{ \large Results for the spherically symmetric harmonic }
\centerline{\large oscillator trap}

\vskip 0.60in
\centerline{
\begin{tabular}{||c|c||}
\hline\hline
 & \\
Size of the Condensate & $ R_{0} = L \left(\frac{15 a N}{L}\right)^{1/5}
$ \\
                  &          \\ \hline
        & \\
Chemical Potential & $\mu = \frac{\hbar \omega}{2} 
\left(\frac{15 a N}{L} \right)^{2/5} $ \\
    &   \\ \hline
    &  \\
Condensate Density & $\phi^{2}(R) =
\frac{R_{0}^{2}}{8 \pi a L^{4}} \left[ 1 - (R/R_{0})^{2} \right]
$ \\
  & \\ \hline
  & \\
Local Coherence Length & 
$\lambda_{c}(R) = \frac{L^{2}}{\sqrt{R_{0}^{2} - R^{2}}}$ \\
      & \\ \hline 
 & \\
Local Depletion  & $ d(R) = \frac{2 \sqrt{2}}{ 3 \pi} 
\frac{a R_{0}}{L^{2}}
\sqrt{1 - \left(R/R_{0} \right)^{2}}$ \\
  &  \\ \hline
  &  \\
Error due to neglecting the Laplacian & 
$ e_{L} (R) = \left( \frac{L}{R_{0}} \right) ^{4}
\frac{ \left[ 3  - 2 (R/R_{0})^{2} \right] }{(1 - (R/R_{0})^{2})^{3}} \; \; 
$ \\
       & \\ \hline
    & \\
Error due the inhomogeneity & 
$e_{i} (R) = \frac{1}{2} \left( \frac{L}{R_{0}} \right) ^{4}
\frac{(R/R_{0})^{2}}{(1 - (R/R_{0})^{2})^{3}} \; \; 
$ \\
      & \\ \hline \hline
\end{tabular}
}

\end{document}